\documentclass[10pt,english,floatfix,superscriptaddress,aps,prd,preprint,nofootinbib]{revtex4}
\usepackage{amsmath}
\usepackage{amssymb}
\usepackage{amsbsy}
\usepackage[a4paper, margin=1.8cm]{geometry}
\usepackage{amsfonts}
\usepackage{amsopn}
\usepackage{amstext}
\usepackage{graphicx}
\usepackage{amssymb}
\usepackage{amsfonts}
\usepackage{amsmath}
\usepackage{graphicx}
\usepackage[english]{babel}
\usepackage{color}
\usepackage{xcolor}
\usepackage{slashed}
\usepackage{esint}
\usepackage[dvips]{epsfig}
\usepackage[dvips]{graphicx}
\usepackage{float}
\usepackage{units}
\usepackage{textcomp}
\usepackage{placeins}
\usepackage{hyperref}             
\hypersetup{
    colorlinks=true,              
    breaklinks=true,              
    citecolor=blue,               
    linkcolor=[rgb]{0,0.5,0.9},   
    urlcolor=red,                 
    filecolor=green               
}

\usepackage{hyperref}
\usepackage{slashed}

\newcommand{\ie}{\begin{equation}}
\newcommand{\fe}{\end{equation}}
 \newcommand{\bq}{\begin{equation}}
 \newcommand{\eq}{\end{equation}}
 \newcommand{\bqn}{\begin{eqnarray}}
 \newcommand{\eqn}{\end{eqnarray}}
 \newcommand{\nb}{\nonumber}
 \newcommand{\lb}{\label}

\begin{document}

\title{Gravitational waves in a minimal gravitational SME}



\author{A. A. Ara\'{u}jo Filho}
\email{dilto@fisica.ufc.br}
\affiliation{Departamento de Física, Universidade Federal da Paraíba, Caixa Postal 5008, 58051--970, João Pessoa, Paraíba,  Brazil.}
\affiliation{Departamento de Física, Universidade Federal de Campina Grande Caixa Postal 10071, 58429-900 Campina Grande, Paraíba, Brazil.}
\affiliation{Center for Theoretical Physics, Khazar University, 41 Mehseti Street, Baku, AZ-1096, Azerbaijan.}
\author{N. Heidari}
\email{heidari.n@gmail.com}

\affiliation{Departamento de Física, Universidade Federal de Campina Grande Caixa Postal 10071, 58429-900 Campina Grande, Paraíba, Brazil.}
\affiliation{Center for Theoretical Physics, Khazar University, 41 Mehseti Street, Baku, AZ-1096, Azerbaijan.}
\affiliation{School of Physics, Damghan University, Damghan, 3671641167, Iran.}


\author{Iarley P. Lobo}
\email{lobofisica@gmail.com}

\affiliation{Departamento de Física, Universidade Federal de Campina Grande Caixa Postal 10071, 58429-900 Campina Grande, Paraíba, Brazil.}
\affiliation{Department of Chemistry and Physics, Federal University of Para\'iba, Rodovia BR 079 - km 12, 58397-000 Areia-PB,  Brazil.}

\date{\today}

\begin{abstract}

In this work, we investigate the generation and propagation of gravitational waves within a minimal gravitational SME (Standard Model Extension). Starting from the modified graviton dispersion relation derived in the linearized gravity sector, we analyze the polarization properties of gravitational waves in the transverse--traceless tensor sector. We then construct the retarded Green function associated with the Lorentz--violating wave operator, explicitly verifying the causal structure of the theory and identifying the modified propagation speeds of the tensorial modes.
In addition, we study the source--induced emission of gravitational waves from a binary black--hole system. We show that the gravitational waveform preserves the standard quadrupolar amplitude and polarization structure, while Lorentz--violating effects enter exclusively through a modification of the retarded time. As a result, the spatial components of the metric perturbation $h_{ij}(t,r)$ acquire a phase shift determined by the SME coefficients. Finally, we estimate phenomenological bounds to the model under consideration. 

\end{abstract}


\maketitle
\tableofcontents

\section{Introduction}

The well--known General Relativity has long provided the standard geometric description of gravitation and has been repeatedly confronted with experimental tests ranging from laboratory scales to astrophysical and cosmological observations, all of which have confirmed its predictions with high accuracy \cite{Manchester:2015mda,Will:1997bb,Hoyle:2000cv,Jain:2010ka,Clifton:2011jh,Koyama:2015vza,Stairs:2003eg,Wex:2014nva,Kramer:2016kwa,Berti:2015itd}. Nevertheless, this theoretical framework remains incomplete. At a fundamental level, it predicts spacetime singularities and resists a consistent formulation at the quantum level, while at large scales it leaves unexplained the empirical evidence commonly attributed to dark matter and dark energy.
Motivated by these features, a variety of approaches aimed at describing gravity beyond the classical regime have been developed. Proposals inspired by quantum gravity, such as string--based models \cite{Kostelecky:1988zi,Kostelecky:1991ak}, loop quantum gravity \cite{Gambini:1998it}, and higher--dimensional brane constructions \cite{Burgess:2002tb}, often depart from one of the central assumptions of GR: the exact validity of Lorentz and diffeomorphism symmetries. In several of these scenarios, these symmetries are not imposed a priori but may instead emerge dynamically or be broken through spontaneous mechanisms in the gravitational sector.

Possible departures from Lorentz and diffeomorphism symmetries in gravitation are most naturally explored within an effective field--theoretic setting. In this spirit, the SME (Standard Model Extension) was formulated as a general framework capable of parametrizing symmetry--breaking effects in a systematic and model-independent manner \cite{Colladay:1996iz, Kostelecky:2003fs}. Rather than modifying the underlying geometric structure directly, the SME incorporates all admissible Lorentz-- and diffeomorphism--violating operators into the Lagrangian, organized according to their mass dimension and symmetry properties. Historically, this framework has been extensively employed to test Lorentz symmetry in the matter sector, yielding a huge class of experimental and observational constraints. When gravitational interactions are included, the SME has enabled precision tests using a wide variety of observational channels, such as lunar laser ranging experiments \cite{Bourgoin:2017fpo,Battat:2007uh}, atom-interferometric measurements \cite{Muller:2007es}, ultra-high-energy cosmic rays \cite{Kostelecky:2015dpa}, pulsar timing observations \cite{Shao:2014oha,Shao:2019tle,Shao:2014bfa,Shao:2019cyt,Jennings:2015vma,Shao:2018vul}, analyses of planetary ephemerides \cite{Hees:2015mga}, and superconducting gravimeter data \cite{Flowers:2016ctv}. In most of these studies, the primary emphasis has been placed on how Lorentz violation manifests through couplings between gravity and matter fields \cite{Kostelecky:2010ze}.
A different perspective emerges when attention is restricted to the gravitational sector itself. In this work, Lorentz-- and diffeomorphism--violating effects are examined directly at the level of gravitational wave propagation, treating the metric perturbations within a linearized regime \cite{Bailey:2006fd,Kostelecky:2017zob,Ferrari:2006gs}.

When gravity is treated as a perturbative field around flat spacetime, Lorentz--violating corrections appear in the quadratic sector of the action through operators of different mass dimensions. These contributions can be organized according to the operator dimension $d$ and separated based on their behavior under linearized coordinate transformations. In particular, invariance under the gauge transformation
$h_{\mu\nu} \rightarrow h_{\mu\nu} + \partial_\mu \xi_\nu + \partial_\nu \xi_\mu$,
with $h_{\mu\nu}$ denoting metric fluctuations about Minkowski spacetime and $\xi_\mu$ an infinitesimal vector field, serves as the criterion distinguishing diffeomorphism--preserving operators from those that explicitly break this symmetry.
Within this classification, operators that respect diffeomorphism invariance arise only at mass dimension \footnote{Here, it is worth mentioning that, in the context of the SME, the designation \emph{minimal} refers to the restriction to power--counting renormalizable operators, namely those with mass dimension $d\leq 4$. In the gravitational sector, and in particular within the linearized framework adopted in this work, this restriction amounts to retaining only the $d=4$ Lorentz--violating operators in the quadratic action for metric perturbations, while excluding higher--dimension ($d>4$) nonminimal contributions.} $d \geq 4$ \cite{Kostelecky:2017zob,Kostelecky:2016kfm}. The impact of such terms on gravitational--wave propagation and the corresponding observational bounds has been examined in detail in numerous studies, including Refs.~\cite{Kostelecky:2016kfm,Hou:2024xbv,Dong:2023nau,Zhu:2023rrx,Mewes:2019dhj,Kostelecky:2017zob,Arzano:2016twc,Kostelecky:2008ts,Haegel:2022ymk,ONeal-Ault:2021uwu}. By contrast, if diffeomorphism symmetry is relaxed at the linearized level, Lorentz--violating operators of lower dimensionality become admissible, allowing contributions with $d=2$ and $d=3$ \cite{Kostelecky:2017zob}.

A complete catalog of all Lorentz--violating operators quadratic in metric perturbations—covering both cases that preserve linearized diffeomorphism symmetry and those that do not—has been established in Ref.~\cite{Kostelecky:2017zob}. The presence of these operators generically reshapes the dispersion properties of gravitational waves, giving rise to phenomena such as direction--dependent propagation, polarization-dependent speeds, and frequency--dependent delays, all of which leave imprints on the observed waveforms.
For scenarios in which diffeomorphism symmetry is maintained, the corresponding modifications to gravitational--wave signals have been explicitly worked out in Ref.~\cite{Mewes:2019dhj}. Those results allow direct confrontation with observational data through Bayesian analyses, enabling bounds to be placed on the relevant Lorentz--violating coefficients within the linearized gravitational sector of the SME. Complementary to this line of work, a parametrized description of symmetry--breaking effects in gravitational--wave propagation has been developed in Ref.~\cite{Zhu:2023rrx}. This formalism has subsequently been applied to assess Lorentz--violating signatures in primordial gravitational waves, as demonstrated in Ref.~\cite{Li:2024fxy}.

In this work, we analyze gravitational wave dynamics within the minimal gravitational SME, emphasizing Lorentz--violating modifications to graviton propagation. Starting from the modified dispersion relation in linearized gravity, we examine the polarization properties of transverse–traceless tensor modes. The retarded Green function of the Lorentz--violating wave operator is derived to clarify the causal structure and the modified propagation speeds. We then study gravitational radiation from a binary black hole system, showing that the waveform preserves the standard quadrupolar amplitude and polarization structure, with Lorentz--violating effects entering only through a shift in the retarded time. This induces a phase modification in the spatial metric perturbations $h_{ij}(t,r)$. Finally, phenomenological constraints on the Lorentz--violating parameters are estimated.


\section{Overview of the Model}

This section introduces the description of gravitational waves in the linearized gravity sector of the SME. In this framework, symmetry--breaking operators modify the dynamics of tensor perturbations, leading to a Lorentz- and diffeomorphism--violating dispersion relation for gravitational waves. The analysis is based on the quadratic action for metric fluctuations around Minkowski spacetime, which encodes the deviations from standard graviton propagation \cite{Kostelecky:2016kfm,Wang:2025fhw}
\bqn \lb{calL}
\mathcal{L} &=& \frac{1}{4}\epsilon^{\mu \rho \alpha \kappa} \epsilon^{\nu \sigma \beta \lambda} \eta_{\kappa \lambda} h_{\mu\nu} \partial_\alpha \partial_\beta h_{\rho \sigma} \nb + \frac{1}{4}h_{\mu\nu} \sum_{{\cal K}, d}  \hat{\cal K}^{(d)\mu\nu\rho\sigma}h_{\rho \sigma}.
\eqn
The gravitational field is treated as a perturbation around flat spacetime by decomposing the metric as $g_{\mu\nu}=\eta_{\mu\nu}+h_{\mu\nu}$, where $\eta_{\mu\nu}$ denotes the Minkowski background. In this formulation, $\epsilon^{\mu\rho\alpha\kappa}$ corresponds to the totally antisymmetric Levi--Civita tensor. The quadratic contribution reproduces the standard Einstein--Hilbert dynamics for linearized gravity, whereas additional terms introduce departures from Lorentz and diffeomorphism invariance. These corrections are encoded through differential operators $\hat{\cal K}^{(d)\mu\nu\rho\sigma}$, constructed by contracting constant coefficients ${\cal K}^{(d)\mu\nu\rho\sigma\alpha_1\ldots\alpha_{d-2}}$ with $d-2$ spacetime derivatives $\partial_{\alpha_1}\cdots\partial_{\alpha_{d-2}}$
\bqn
\hat{\cal K}^{(d)\mu\nu\rho\sigma} = {\cal K}^{(d)\mu\nu\rho\sigma \alpha _1 \alpha _2 \cdots \alpha _{d-2}}{\partial_{\alpha _1 } \partial_{\alpha _2} \cdots \partial_{\alpha _{d-2}}}.
\eqn

The tensors ${\cal K}^{(d)\mu\nu\rho\sigma\alpha_1\ldots\alpha_{d-2}}$ carry mass dimension $4-d$ and are treated as small, spacetime-independent parameters over the physical scales of interest. Their contribution to the linearized field equations arises only when the associated operator $\hat{\cal K}^{(d)\mu\nu\rho\sigma}$ is not symmetric under the exchange of the index pairs $(\mu\nu)$ and $(\rho\sigma)$, namely when
$K^{(d)(\mu\nu)(\rho\sigma)} \neq \pm K^{(d)(\rho\sigma)(\mu\nu)}$,
with the upper (lower) sign corresponding to odd (even) values of $d$ \cite{Kostelecky:2016kfm,Wang:2025fhw}
\bqn
\hat{\cal K}^{(d)\mu\nu\rho\sigma} = {\hat s}^{(d)\mu\rho\nu\sigma}+{\hat q}^{(d)\mu\rho\nu\sigma}+{\hat k}^{(d)\mu\rho\nu\sigma}.
\eqn
The Lorentz--violating operators separate naturally into three families, distinguished by how their indices transform under permutations. One class, denoted by $\hat{s}^{(d)\mu\rho\nu\sigma}$, carries antisymmetry within each index pair $(\mu\rho)$ and $(\nu\sigma)$. A second class, $\hat{q}^{(d)\mu\rho\nu\sigma}$, combines antisymmetry in $(\mu\rho)$ with symmetry in $(\nu\sigma)$, while the remaining operators $\hat{k}^{(d)\mu\rho\nu\sigma}$ are fully symmetric under all index exchanges. Each family admits a further decomposition into irreducible components, allowing a characterization of their impact on gravitational--wave propagation. This procedure yields fourteen independent operator classes, as summarized in Table I of Refs.~\cite{Kostelecky:2016kfm,Wang:2025fhw}.

Among these, the $\hat{s}$--type operators $\hat{s}^{(d)\mu\rho\nu\sigma}$ are CPT--even, meaning they remain invariant under the combined action of charge conjugation, parity, and time reversal. Their structure permits a decomposition into three distinct irreducible contributions
\bqn
\hat{s}^{{(d)\mu\rho\nu\sigma}} = \hat{s}^{(d, 0)\mu\rho\nu\sigma} + \hat{s}^{(d, 1)\mu\rho\nu\sigma} +  \hat{s}^{(d, 2)\mu\rho\nu\sigma},
\eqn
where
\bqn\lb{sd}
{\hat{s}}^{(d,0)\mu \rho \nu \sigma} &=& {s}^{(d,0) \mu \rho \alpha_{1} \nu \sigma \alpha_{2} \alpha_3 \ldots \alpha_{d-2}} \partial_{\alpha_{1}} \ldots \partial_{\alpha_{d-2}}, \lb{sd0}\nb\\
{\hat{s}}^{(d,1)\mu \rho \nu \sigma} &=& {s}^{(d,1) \mu \rho \nu \sigma \alpha_{1}  \ldots \alpha_{d-2}} \partial_{\alpha_{1}} \ldots \partial_{\alpha_{d-2}}, \lb{sd1}\nb\\
{\hat{s}}^{(d,2)\mu \rho \nu \sigma} &=& {s}^{(d,2) \mu \rho \alpha_{1} \nu \sigma \alpha_{2} \alpha_3 \ldots \alpha_{d-2}} \partial_{\alpha_{1}} \ldots \partial_{\alpha_{d-2}}. \lb{sd2}
\eqn

On the other hand, the operators of $\hat{q}$ type, denoted by $\hat{q}^{(d)\mu\rho\nu\sigma}$, violate CPT symmetry and therefore reverse sign under the combined action of charge conjugation, parity, and time reversal. Their algebraic structure allows them to be separated into six independent irreducible components
\bqn
\hat{q}^{(d)\mu\rho\nu\sigma} &=& \hat{q}^{(d, 0)\mu\rho\nu\sigma}+\hat{q}^{(d, 1)\mu\rho\nu\sigma}+\hat{q}^{(d, 2)\mu\rho\nu\sigma}\nb +\hat{q}^{(d, 3)\mu\rho\nu\sigma}+\hat{q}^{(d, 4)\mu\rho\nu\sigma}+\hat{q}^{(d, 5)\mu\rho\nu\sigma},\nb\\
\eqn
with
\bqn\lb{qd}
\hat{q}^{(d, 0)\mu\rho\nu\sigma}&=&\hat{q}^{(d, 0)\mu\rho\alpha_1 \nu \alpha_2 \sigma \alpha_3 \alpha_4 \ldots \alpha_{d-2}} \partial_{\alpha_{1}} \ldots \partial_{\alpha_{d-2}}, \nb\\
\hat{q}^{(d, 1)\mu\rho\nu\sigma}&=&\hat{q}^{(d, 1)\mu\rho \nu  \sigma \alpha_1 \alpha_2 \ldots \alpha_{d-2}} \partial_{\alpha_{1}} \ldots \partial_{\alpha_{d-2}}, \nb\\
\hat{q}^{(d, 2)\mu\rho\nu\sigma}&=&\hat{q}^{(d, 2)\mu\rho \nu \alpha_1 \sigma \alpha_2  \ldots \alpha_{d-2}} \partial_{\alpha_{1}} \ldots \partial_{\alpha_{d-2}}, \nb\\
\hat{q}^{(d, 3)\mu\rho\nu\sigma}&=&\hat{q}^{(d, 3)\mu\rho\alpha_1 \nu  \sigma \alpha_2 \ldots \alpha_{d-2}} \partial_{\alpha_{1}} \ldots \partial_{\alpha_{d-2}}, \nb\\
\hat{q}^{(d, 4)\mu\rho\nu\sigma}&=&\hat{q}^{(d, 4)\mu\rho\nu \alpha_1 \sigma \alpha_2 \alpha_3 \alpha_4 \ldots \alpha_{d-2}} \partial_{\alpha_{1}} \ldots \partial_{\alpha_{d-2}}, \nb\\
\hat{q}^{(d, 5)\mu\rho\nu\sigma}&=&\hat{q}^{(d, 5)\mu\rho \alpha_1 \nu \alpha_2 \sigma \alpha_3 \alpha_4 \ldots \alpha_{d-2}} \partial_{\alpha_{1}} \ldots \partial_{\alpha_{d-2}}.\nb\\
\eqn
Operators belonging to the $\hat{k}$ sector preserve CPT symmetry and admit a decomposition into five distinct irreducible contributions
\bqn
\hat{k}^{(d)\mu\nu\rho\sigma} &=& \hat{k}^{(d, 0)\mu\nu\rho\sigma} +\hat{k}^{(d, 1)\mu\nu\rho\sigma} +\hat{k}^{(d, 2)\mu\nu\rho\sigma} \nb +\hat{k}^{(d, 3)\mu\nu\rho\sigma} +\hat{k}^{(d, 4)\mu\nu\rho\sigma} ,
\eqn
so that
\bqn\lb{kd}
\hat{k}^{(d, 0)\mu\nu\rho\sigma} &=& \hat{k}^{(d, 0)\mu \alpha_1 \nu \alpha_2 \rho \alpha_3 \sigma \alpha_4 \alpha_5 \ldots \alpha_{d-2}}  \partial_{\alpha_{1}} \ldots \partial_{\alpha_{d-2}}, \nb\\
\hat{k}^{(d, 1)\mu\nu\rho\sigma} &=& \hat{k}^{(d, 1)\mu \nu  \rho \sigma \alpha_1 \ldots \alpha_{d-2}}  \partial_{\alpha_{1}} \ldots \partial_{\alpha_{d-2}}, \nb\\
\hat{k}^{(d, 2)\mu\nu\rho\sigma} &=& \hat{k}^{(d, 2)\mu \alpha_1 \nu  \rho  \sigma \alpha_1 \alpha_2 \ldots \alpha_{d-2}}  \partial_{\alpha_{1}} \ldots \partial_{\alpha_{d-2}}, \nb\\
\hat{k}^{(d, 3)\mu\nu\rho\sigma} &=& \hat{k}^{(d, 3)\mu \alpha_1 \nu \alpha_2 \rho  \sigma \alpha_3 \alpha_5 \ldots \alpha_{d-2}}  \partial_{\alpha_{1}} \ldots \partial_{\alpha_{d-2}}, \nb\\
\hat{k}^{(d, 4)\mu\nu\rho\sigma} &=& \hat{k}^{(d, 4)\mu \alpha_1 \nu \alpha_2 \rho \alpha_3 \sigma \alpha_4 \alpha_5 \ldots \alpha_{d-2}}  \partial_{\alpha_{1}} \ldots \partial_{\alpha_{d-2}}. \nb\\
\eqn
Furthermore, the Lorentz--violating contributions in linearized gravity can be organized into fourteen independent sets of coefficients, whose defining properties are listed in Table I of Refs.~\cite{Kostelecky:2016kfm,Wang:2025fhw}. These ones collectively determine all admissible departures from standard graviton propagation at the linear level, encoding the full range of observable effects in gravitational wave dynamics.

Imposing invariance under infinitesimal coordinate transformations places strong restrictions on this structure. Requiring the quadratic action $S\sim\int \mathrm{d}^4x\,\mathcal{L}$ to remain unchanged under
$h_{\mu\nu}\rightarrow h_{\mu\nu}+\partial_\mu\xi_\nu+\partial_\nu\xi_\mu$
forces the Lorentz--violating operator to satisfy
$\hat{\cal K}^{(d)(\mu\nu)(\rho\sigma)}\partial_\nu=\pm\hat{\cal K}^{(d)(\rho\sigma)(\mu\nu)}\partial_\nu$.
Once this condition is enforced, the allowed operator content collapses to three distinct families, denoted by ${\hat s}^{(d,0)\mu\rho\nu\sigma}$, ${\hat q}^{(d,0)\mu\rho\nu\sigma}$, and ${\hat k}^{(d,0)\mu\rho\nu\sigma}$ \cite{Kostelecky:2016kfm,Wang:2025fhw}. Preservation of diffeomorphism symmetry further restricts Lorentz--violating effects to operators with mass dimension $d\geq4$, whereas abandoning this symmetry permits contributions already at $d\geq2$.

The field equations governing gravitational--wave propagation follow from the quadratic action $S\sim\int \mathrm{d}^4x\,\mathcal{L}$ by performing a variation with respect to the metric perturbation $h_{\mu\nu}$. Using the Lagrangian density defined in Eq.~(\ref{calL}), this procedure leads to
\bqn\lb{eom0}
\frac{1}{2}\eta_{\rho \sigma} \epsilon^{\mu\rho\alpha \kappa} \epsilon^{\nu\sigma \beta \lambda} \partial_\alpha \partial_\beta h_{\kappa \lambda} - \delta M^{\mu\nu \rho\sigma} h_{\rho \sigma}=0,
\eqn
in which the tensor operators are written as follows
\bqn
\delta M^{\mu\nu\rho \sigma} &=& - \frac{1}{4} \left(\hat{s}^{\mu\rho \nu \sigma} + \hat{s}^{\mu \sigma \nu \rho}\right) - \frac{1}{2} \hat{k}^{\mu \nu \rho \sigma} -\frac{1}{8} \left(\hat{q}^{\mu \rho \nu \sigma} + \hat{q}^{\nu \rho \mu \sigma} +\hat{q}^{\mu \sigma \nu \rho} + \hat{q}^{\nu \sigma \mu \rho}\right).\nb\\
\eqn
where
\bqn
\hat{s}^{\mu\rho \nu \sigma}  = \sum_{d} \hat{s}^{(d)\mu\rho \nu \sigma}, \\
\hat{q}^{\mu\rho \nu \sigma}  = \sum_{d} \hat{q}^{(d)\mu\rho \nu \sigma}, \\
\hat{k}^{\mu\rho \nu \sigma}  = \sum_{d} \hat{k}^{(d)\mu\rho \nu \sigma}.
\eqn

Within general relativity, gravitational radiation is carried exclusively by two transverse and traceless tensor degrees of freedom. Once Lorentz-- or diffeomorphism--breaking effects are incorporated, this minimal structure need not be preserved, and additional propagating components may arise. Depending on the underlying symmetry breaking, the metric perturbation can support extra scalar and vector polarizations, supplementing the usual tensor modes. In Lorentz--violating linearized gravity, these non--tensorial excitations can be sourced indirectly by the tensor sector itself, as demonstrated in Ref.~\cite{Hou:2024xbv,Wang:2025fhw}.

Despite this theoretical possibility, current gravitational--wave observations provide no evidence for departures from the standard tensorial polarization pattern. Measurements reported by the LIGO--Virgo--KAGRA Collaboration remain consistent with the presence of only two tensor modes, with no statistically meaningful indication of scalar or vector contributions \cite{KAGRA:2021vkt}. Even if such additional modes were present, their amplitudes would be expected to remain suppressed relative to the dominant tensor signal.

Motivated by both observational results and analytical simplicity, the present analysis is restricted to the transverse--traceless tensor sector. Treating any extra polarizations as negligible, and adopting an approach analogous to that of Ref.~\cite{Kostelecky:2016kfm}, we examine how Lorentz-- and diffeomorphism--violating operators modify the propagation of these two tensor modes alone.


\section{Polarization properties of gravitational waves}

In the case of operators with mass dimension $d=4$, the resulting dispersion relation assumes a particularly simple form \cite{Kostelecky:2016kfm}
\begin{equation}
\label{mdr5}
p^0=\Big(1-\varsigma_0 \pm \sqrt{\varsigma_1^2+\varsigma_2^2+\varsigma_3^2}\Big)\,|\vec p| .
\end{equation}
Here, we have the wave four--momentum is denoted by $p^\mu=(p^0,\vec p)$. The functions $\varsigma_i$ encode the Lorentz--violating effects and arise as momentum--dependent combinations of gauge--invariant operators appearing in the quadratic action for gravitational perturbations. Their explicit definitions are given in Ref.~\cite{Kostelecky:2016kfm}
\begin{align}
\varsigma_0 &=
\frac{1}{4p^2}
\left(
-\,\hat s^{\mu\nu}{}_{\mu\nu}
+\frac{1}{2}\,\hat k^{\mu\nu}{}_{\mu\nu}
\right), \\[1ex]
(\varsigma_1)^2+(\varsigma_2)^2 &=
\frac{1}{8p^4}
\left(
\hat k^{\mu\nu\rho\sigma}\hat k_{\mu\nu\rho\sigma}
-\hat k^{\mu\rho}{}_{\nu\rho}\,\hat k_{\mu\sigma}{}^{\nu\sigma}
+\frac{1}{8}\,\hat k^{\mu\nu}{}_{\mu\nu}\,
\hat k^{\rho\sigma}{}_{\rho\sigma}
\right), \\[1ex]
(\varsigma_3)^2 &=
\frac{1}{16p^4}
\left(
-\frac{1}{2}\,\hat q^{\mu\rho\nu\sigma}\hat q_{\mu\rho\nu\sigma}
-\hat q^{\mu\nu\rho\sigma}\hat q_{\mu\rho\nu\sigma}
+\big(\hat q^{\mu\rho\nu}{}_{\rho}+\hat q^{\nu\rho\mu}{}_{\rho}\big)
\hat q_{\mu\sigma\nu}{}^{\sigma}
\right),
\end{align}
where symbols carrying hats represent the momentum--space forms of the Lorentz--violating operators, obtained by implementing the replacement $\partial_\mu \rightarrow i p_\mu$.

Eq.~\eqref{mdr5} reveals that the modifications to gravitational--wave propagation manifest through three qualitatively different mechanisms due to Lorentz violation. One effect concerns direction dependence: when rotational symmetry is broken, the wave dynamics become anisotropic, with the corresponding behavior determined—within a chosen observer frame—by coefficients carrying spatial components. A second effect involves frequency dependence of the propagation speed. Such dispersive behavior arises whenever momentum--dependent operators are present, implying that only terms with mass dimension $d=4$ permit frequency--independent propagation, governed solely by the operator $\hat{s}^{\mu\rho\nu\sigma}$. A third possibility is the splitting of polarization modes during propagation. The appearance of two separate solution branches in Eq.~\eqref{mdr5} signals birefringent behavior, which can occur exclusively through the operators $\hat{q}^{\mu\rho\nu\sigma}$ and $\hat{k}^{\mu\nu\rho\sigma}$ and therefore only when operators with $d>4$ are involved.

Through straightforward algebra, the dispersion relation admits an equivalent series representation \cite{Kostelecky:2016kfm}
\begin{equation}
\label{kkkk}
\omega
=
\left(1-\mathring{k}^{(4)}_{(I)}\right)\,|\mathbf{p}|
\;\pm\;
\mathring{k}^{(5)}_{(V)}\,\omega^{2}
\;-\;
\mathring{k}^{(6)}_{(I)}\,\omega^{3}
\;\pm\;
\mathring{k}^{(7)}_{(V)}\,\omega^{4}
\;-\;
\mathring{k}^{(8)}_{(I)}\,\omega^{5}
\;\pm\;\ldots
\end{equation}

For simplicity, we restrict our analysis to the case $d=4$, retaining only the coefficient
$\mathring{k}^{(4)}_{(I)}$. While Ref.~\cite{Kostelecky:2016kfm} reports bounds for operators
with $d>4$, the $d=4$ sector remains unconstrained. As shown in the final section, bounds on
$\mathring{k}^{(4)}_{(I)}$ are obtained here from the modified group velocity.

On the other hand, the polarization content of gravitational waves encodes essential information about both their astrophysical origin and the fundamental symmetries governing gravity. In particular, departures from local Lorentz invariance can manifest directly in the polarization structure, making polarization observables a powerful probe of gravitational dynamics and wave propagation \cite{le2017theory,liang2017polarizations,zhang2018velocity,hou2018polarizations,mewes2019signals,liang2022polarizations,Nilsson:2023szw,Bailey:2023lzy}.

In this work, gravitational radiation is described as a perturbation propagating over a flat spacetime background with metric $\eta_{\mu\nu}=\mathrm{diag}(-1,1,1,1)$. Focusing on the tensor sector, the spatial components of the metric perturbation, $h_{ij}(x)$ with $i,j=1,2,3$, satisfy the homogeneous equations of motion appropriate for freely propagating gravitational waves \cite{Wang:2025fhw}
\begin{equation}
\label{EOM_SME_TT}
\big(\partial_t^2-\nabla^2\big)\,h_{ij}(x)
\;+\;2\,\delta M_{ijmn}(\partial)\,h_{mn}(x)=0 ,
\end{equation}
with
\begin{equation}
\partial_t \equiv \frac{\partial}{\partial t},
\qquad
\nabla^2 \equiv \delta^{kl}\partial_k\partial_l.
\nonumber
\end{equation}
Here, derivatives are taken with respect to the flat spacetime background. Lorentz--violating corrections enter through the differential operator $\delta M_{ijmn}(\partial)$, which acts linearly on the tensor perturbation $h_{mn}$. This operator is constructed from fixed background tensors contracted with spacetime derivatives, and its detailed form is determined by the specific Lorentz--violating sector under consideration, potentially incorporating terms of various mass dimensions.

As argued before, the transverse--traceless condition of the metric perturbation is maintained
\begin{equation}
\label{constraints}
h_{ii}=0,
\qquad
\partial_i h_{ij}=0 .
\end{equation}
As a result, non--tensorial modes are excluded, and the dynamics reduce to the two physical tensor degrees of freedom that carry gravitational radiation. This restriction originates from the residual gauge symmetry of the linearized framework and continues to be valid even after Lorentz--violating terms are incorporated.

The polarization characteristics are extracted by decomposing the gravitational perturbations into plane--wave modes written as
\begin{equation}
h_{\mu\nu}(x)=\varepsilon_{\mu\nu}\,e^{i p_\mu x^\mu},
\end{equation}
where the four--momentum is
\begin{equation}
p^\mu=(p^0,0,0,p^3).
\end{equation}
The wave is taken to propagate along the $z$ direction. Also, the allowed polarization configurations are constrained by their transversality with respect to the wave four--momentum
\begin{equation}
\label{transversality}
p^\mu \varepsilon_{\mu\nu}=0 .
\end{equation}
In other words, this condition enforces transversality relative to the propagation direction, leading to
\begin{equation}
\varepsilon_{0\mu}=\varepsilon_{3\mu}=0 .
\end{equation}
Moreover, taking into account traceless condition, we obtain
\begin{equation}
\varepsilon_{11}+\varepsilon_{22}=0 ,
\end{equation}
a general representation of the polarization tensor follows once the transverse–traceless conditions are imposed as follows
\begin{equation}
\varepsilon_{\mu\nu}=
\begin{pmatrix}
0 & 0 & 0 & 0 \\
0 & \varepsilon_{11} & \varepsilon_{12} & 0 \\
0 & \varepsilon_{12} & -\varepsilon_{11} & 0 \\
0 & 0 & 0 & 0
\end{pmatrix}.
\end{equation}

Notice that The polarization tensor contains two independent amplitudes, with $\varepsilon_{11}$ and $\varepsilon_{12}$ spanning the tensor sector and playing roles analogous to the conventional plus and cross modes. Departures from standard gravitational dynamics can arise once Lorentz--violating operators are introduced, altering the propagation law of gravitational waves.


\section{Source induced gravitational wave emission}

Gravitational--wave emission is incorporated by promoting the vacuum graviton equation to a sourced system through the introduction of a tensor current $J_{\mu\nu}$. When this source is coupled to the linearized dynamics described by Eq.~\eqref{EOM_SME_TT}, the resulting field equation takes the form
\begin{equation}
  \big(\partial_t^2-\nabla^2\big)\,h_{ij}(x)
\;+\;2\,\delta M_{ijmn}(\partial)\,h_{mn}(x)= J_{\mu\nu}(x).
\end{equation}

In this manner, the resulting perturbation is given by
\begin{equation}
    h_{ij}(x) = \int \mathrm{d}^4y \; G(x-y) J_{ij}(y).
\end{equation}

The key step is the construction of the operator $G(x-y)$. To facilitate this task and streamline the subsequent analysis, the modified dispersion relation introduced in Eq.~(\ref{kkkk}) is recast into a more convenient form
\begin{equation}
\label{modddss}
\omega=p^{0} = v_{\pm}\,|\vec{p}| \,,
\qquad
v_{\pm} \equiv 1 - \mathring{k}^{(4)}_{(I)} \, .
\end{equation}
The wave operator $\partial_t^{2}-v_{\pm}^{2}\nabla^{2}$ admits the following Green function in momentum space:
\begin{equation}
\tilde{G}_{\pm}(p)
=
\frac{1}{(p^{0})^{2} - v_{\pm}^{2} |\vec{p}|^{2}} \, .
\end{equation}
Causal propagation is enforced by defining the Green function with the appropriate retarded configuration,
\begin{equation}
\tilde{G}^{\mathrm{ret}}_{\pm}(p)
=
\frac{1}{(p^{0})^{2} - v_{\pm}^{2} |\vec{p}|^{2} + i\epsilon p^{0}} \, .
\end{equation}
Notice that this choice enforces causal behavior, ensuring that the Green function has support only for positive times. The corresponding retarded Green function in position space then follows from performing the inverse Fourier transformation.
\begin{equation}
G^{\mathrm{ret}}_{\pm}(t,\vec{x})
=
\int \frac{\mathrm{d}p^{0}}{2\pi}
\int \frac{\mathrm{d}^{3}p}{(2\pi)^{3}}
\,
\frac{e^{-ip^{0}t + i\vec{p}\cdot\vec{x}}}
{(p^{0})^{2} - v_{\pm}^{2} |\vec{p}|^{2} + i\epsilon p^{0}} \, .
\label{GretFourier}
\end{equation}

The time component of the momentum integral is evaluated first by extending $p^{0}$ into the complex plane. Imposing the retarded prescription displaces both singularities below the real axis, so that for positive times the integration contour is closed in the lower half--plane. The resulting expression then follows directly from the residues enclosed by the contour
\begin{equation}
\int \frac{\mathrm{d}p^{0}}{2\pi}
\,
\frac{e^{-ip^{0}t}}
{(p^{0})^{2} - v_{\pm}^{2} p^{2} + i\epsilon p^{0}}
=
\Theta(t)\,
\frac{\sin(v_{\pm} p\, t)}{v_{\pm} p} \, ,
\end{equation}
with $p$ denoting the magnitude of the spatial momentum, $p\equiv|\vec p|$, while $\Theta(t)$ represents the Heaviside step function. Inserting this expression into Eq.~\eqref{GretFourier} then yields
\begin{equation}
G^{\mathrm{ret}}_{\pm}(t,\vec{x})
=
\Theta(t)
\int \frac{\mathrm{d}^{3}p}{(2\pi)^{3}}
\,
e^{i\vec{p}\cdot\vec{x}}
\frac{\sin(v_{\pm} p\, t)}{v_{\pm} \,p} \, .
\end{equation}

The remaining angular dependence is removed by performing the integration in spherical momentum coordinates, which leads to
\begin{equation}
\int \mathrm{d}\Omega_{\hat{p}}\,
e^{i\vec{p}\cdot\vec{x}}
=
4\pi \frac{\sin(p\,r)}{p\,r} \, ,
\qquad r \equiv |\vec{x}| \, .
\end{equation}
The Green function is thereby reduced to a single radial integral
\begin{equation}
G^{\mathrm{ret}}_{\pm}(t,r)
=
\Theta(t)\,
\frac{1}{2\pi^{2} v_{\pm} r}
\int_{0}^{\infty} \mathrm{d}p\;
\sin(pr)\,\sin(v_{\pm} p\, t) \, .
\end{equation}

Using the distributional identity
\begin{equation}
\int_{0}^{\infty} \mathrm{d}p\;
\sin(ap)\,\sin(bp)
=
\frac{\pi}{2}\,\delta(a-b)
\qquad (a,b>0),
\end{equation}
the integral can be evaluated in closed form, leading to
\begin{equation}
G^{\mathrm{ret}}_{\pm}(t,r)
=
\Theta(t)\,
\frac{1}{4\pi v_{\pm} r}\,
\delta(r - v_{\pm} t) \, .
\end{equation}
In addition, we can equivalently rewrite the delta function as
\(
\delta(r - v_{\pm} t)
=
\frac{1}{v_{\pm}}\,
\delta\!\left(t - \frac{r}{v_{\pm}}\right),
\)
the retarded Green's function in coordinate space takes the compact form
\begin{equation}
\label{retardeddd}
G^{\mathrm{ret}}_{\pm}(t,\vec{x})
=
\Theta(t)\,
\frac{1}{4\pi v_{\pm}^{2} r}\,
\delta\!\left(t - \frac{r}{v_{\pm}}\right),
\qquad \text{where} \,\, (r = |\vec{x}|).
\end{equation}
In other words, it is worthy higliting that above expression represents a causal signal propagating at speed $v_{\pm}$ and satisfies
\(
(\partial_{t}^{2} - v_{\pm}^{2}\nabla^{2})\,G^{\mathrm{ret}}_{\pm}
=
\delta(t)\delta^{(3)}(\vec{x})
\),
as expected.

Within linearized general relativity, gravitational radiation observed far from the source is encoded in the spatial components of the metric perturbation, which take the form
\begin{equation}
h_{ij}(t,{\bf r})
=
\frac{2G}{r}\,
\frac{\mathrm{d}^{2} I_{ij}}{\mathrm{d}t^{2}}(t_{r}),
\end{equation}
with $t_r=t-r$ representing the retarded time and $I_{ij}$ denoting the mass quadrupole moment as shown below
\begin{equation}
I_{ij}(t)
=
\int y^{i} y^{j} T^{00}(t,\vec y)\,\mathrm{d}^{3}y .
\end{equation}

Furthermore, when Lorentz--violating corrections are incorporated through the modified dispersion relation in Eq.~(\ref{modddss}), gravitational perturbations propagate according to the causal Green function given in Eq.~(\ref{retardeddd}). In this framework, sources described by
$J_{\mu\nu}=16\pi G,T_{\mu\nu}$
generate metric fluctuations that retain the same formal dependence as in standard General Relativity, with the sole difference that the signal propagation is governed by a shifted retarded time
\begin{equation}
h_{ij}(t,{\bf r})
=
\frac{2G}{r}\,
\frac{\mathrm{d}^{2} I_{ij}}{\mathrm{d}t^{2}}\!\left(t_{r}^{(\pm)}\right),
\end{equation}
in which
\begin{equation}
t_{r}^{(\pm)}
=
t - \frac{r}{v_{\pm}} .
\end{equation}
In other words, causal propagation is preserved, but the phase velocity is reduced to $v_{\pm}$, so the signal reaches the observer later than in standard relativity for $v_{\pm}<1$.

\begin{figure}
    \centering
    \includegraphics[scale=0.7]{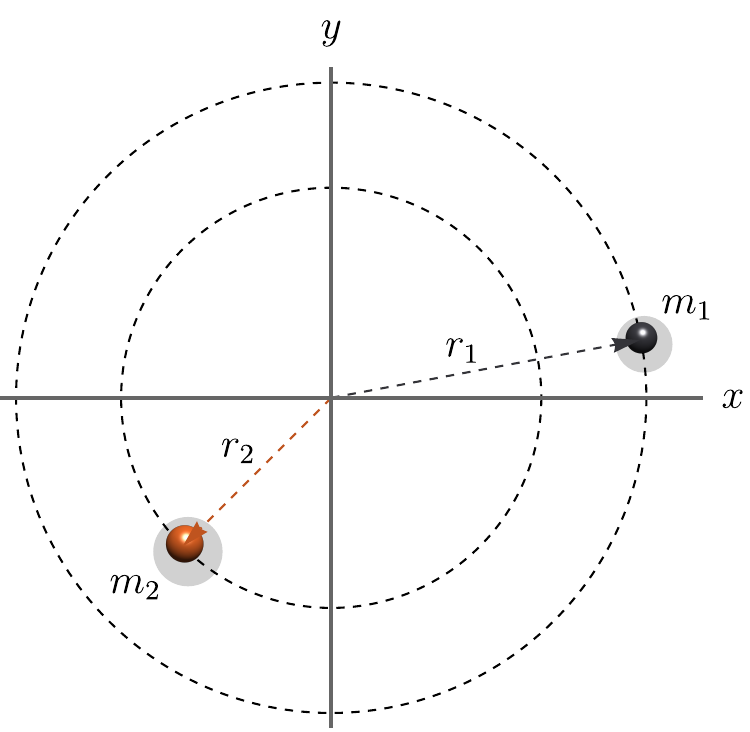}
    \caption{Binary black-hole configuration in the center-of-mass frame. Two compact masses, $m_1$ and $m_2$, orbit within the $xy$ plane with orbital radii $r_1$ and $r_2$.}
    \label{binary}
\end{figure}

We next turn to a concrete application by examining how the Lorentz--violating modifications manifest in a binary black hole configuration, illustrated in Fig.~\ref{binary}. The system consists of two compact objects with masses $m_1$ and $m_2$ executing orbital motion confined to the $xy$ plane. Their trajectories are described relative to the center of mass frame, with the individual orbital radii denoted by $r_1$ and $r_2$, respectively.

The gravitational source is described by an idealized matter distribution consisting of two point masses moving within the orbital plane. Their contribution to the gravitational field is encoded in the corresponding stress--energy tensor, which takes the form
\begin{equation}
T_{00}
=
\delta(z)\Big[m_{1}\,\delta(x-x_{1})\delta(y-y_{1})
+
m_{2}\,\delta(x-x_{2})\delta(y-y_{2})\Big].
\end{equation}

Circular motion about the center-of-mass frame is assumed for both components of the binary, and their dynamics are described by the relations
\begin{align}\label{eqmovbin}
x_{1}(t)=\frac{m_{2}l_{0}}{M}\cos(\omega t),
&\qquad
y_{1}(t)=\frac{m_{2}l_{0}}{M}\sin(\omega t), \nonumber\\
x_{2}(t)=-\frac{m_{1}l_{0}}{M}\cos(\omega t),
&\qquad
y_{2}(t)=-\frac{m_{1}l_{0}}{M}\sin(\omega t),
\end{align}
Here, the binary is characterized by the total mass $M=m_1+m_2$, while the relative separation between the components is fixed by $l_0=r_1+r_2$. The orbital motion proceeds with angular frequency $\omega$. Under these conditions, the structure of the mass quadrupole simplifies considerably, leaving only three components different from zero
\begin{align}\label{compI}
I_{xx}(t)
&=
\frac{\mu}{2}l_{0}^{2}\big[1+\cos(2\omega t)\big], \nonumber\\
I_{yy}(t)
&=
\frac{\mu}{2}l_{0}^{2}\big[1-\cos(2\omega t)\big], \nonumber\\
I_{xy}(t)
=
I_{yx}(t)
&=
\frac{\mu}{2}l_{0}^{2}\sin(2\omega t),
\end{align}
with the reduced mass of the system being defined as $\mu=m_1m_2/(m_1+m_2)$. The causal structure established through the retarded Green function implies that the emitted gravitational signal retains the familiar quadrupole form. The only departure from the standard expression arises from the evaluation time: all time-dependent quantities are to be computed at a shifted retarded time. As a result, the spatial components of the metric perturbation in the radiation zone follow directly by replacing $t$ with $t_r^{(\pm)}$ in the second time derivative of the quadrupole tensor
\begin{equation}
h_{ij}(t,{\bf r})
=
\frac{2G}{r}\,
\frac{\mathrm{d}^{2}I_{ij}}{\mathrm{d}t^{2}}\!\left(t_{r}^{(\pm)}\right).
\nonumber
\end{equation}
As it is clearly seen, the effect of the altered propagation speed $v_{\pm}$ therefore manifests solely as a phase modification in the gravitational signal, without affecting either its amplitude or its tensorial pattern. Substituting the explicit quadrupole expressions from Eq.~\eqref{compI} into the second time derivatives and evaluating them at the shifted retarded time $t_r^{(\pm)}$ leads directly to the nonzero components of the metric perturbation observed in the radiation zone
\begin{align}
h_{xx}^{(\pm)}(t,r)
&=
-\frac{4G\,\mu\,l_{0}^{2}\,\omega^{2}}{r}\,
\cos\!\Big(2\omega\,t_{r}^{(\pm)}\Big), \label{hxx} \\
h_{yy}^{(\pm)}(t,r)
&=
+\frac{4G\,\mu\,l_{0}^{2}\,\omega^{2}}{r}\,
\cos\!\Big(2\omega\,t_{r}^{(\pm)}\Big), \label{hyy} \\
h_{xy}^{(\pm)}(t,r)
=
h_{yx}^{(\pm)}(t,r)
&=
-\frac{4G\,\mu\,l_{0}^{2}\,\omega^{2}}{r}\,
\sin\!\Big(2\omega\,t_{r}^{(\pm)}\Big). \label{hxy}
\end{align}
In general lines, the resulting waveforms therefore maintain the conventional quadrupole amplitude and tensor polarization content. Lorentz-violating contributions do not alter these features but instead appear solely as a modification of the oscillation phase, reflecting the change in the propagation speed $v_{\pm}$.

\begin{figure}
    \centering
    \includegraphics[scale=0.5]{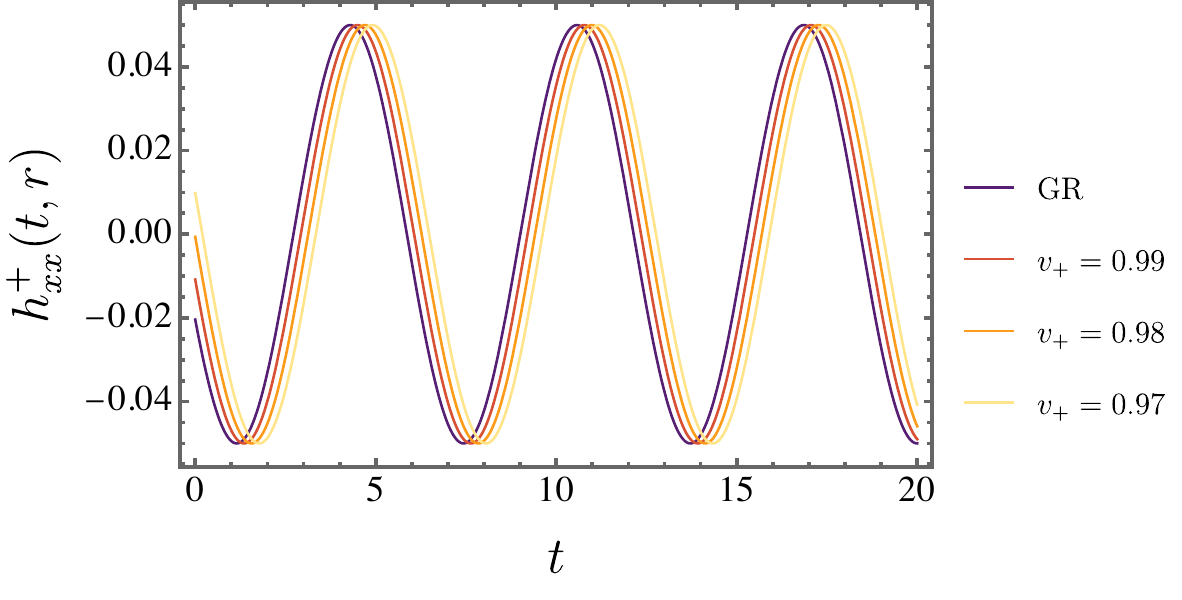}
     \includegraphics[scale=0.5]{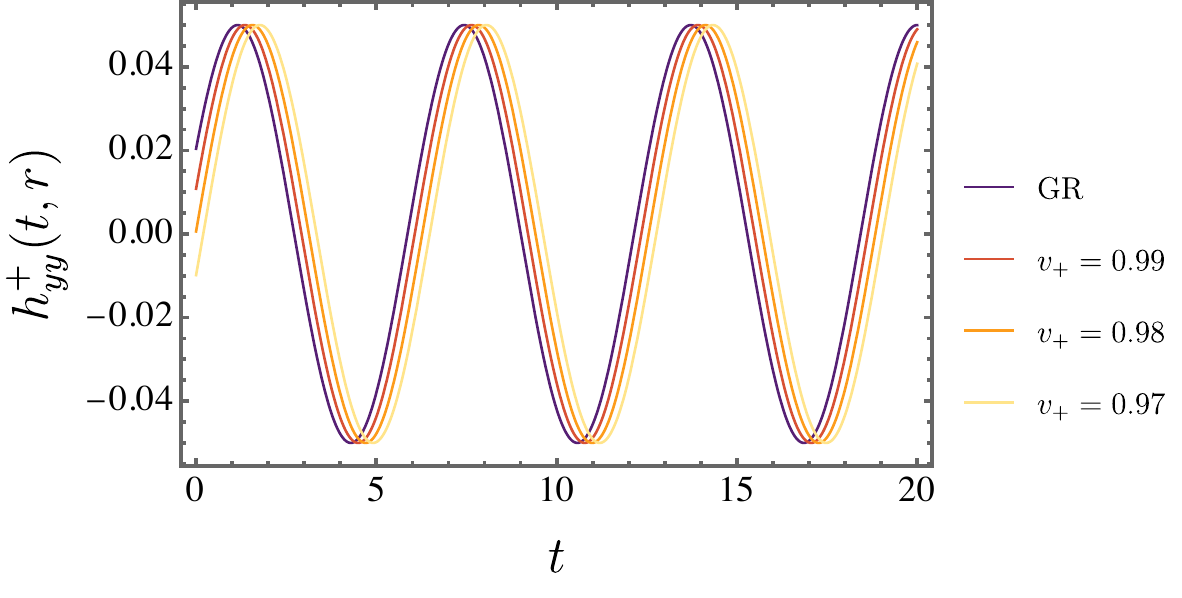}
    \caption{The spatial components $h_{xx}^{(\pm)}(t,r)$ (up panel) and $h_{yy}^{(\pm)}(t,r)$ (down panel) are shown as functions of time $t$ for several representative values of the propagation speed $v_{\pm}$. For reference, the standard general relativity result, corresponding to $v_{\pm}=1$, is also displayed for comparison.}
    \label{hxx_waveform}
\end{figure}

\begin{figure}
    \centering
    \includegraphics[scale=0.5]{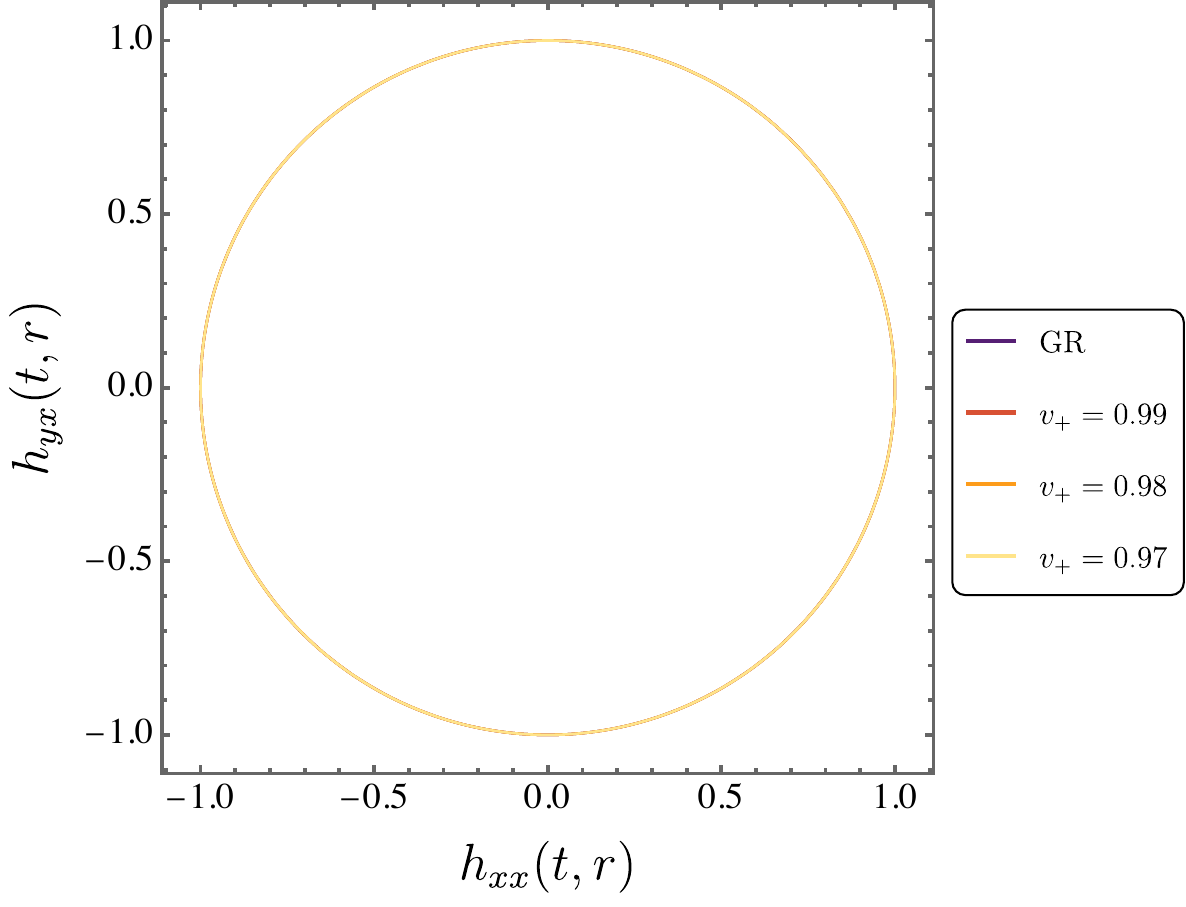}
    \caption{The joint evolution of the components $h_{yx}^{(\pm)}(t,r)$ and $h_{xx}^{(\pm)}(t,r)$ is illustrated through parametric curves evaluated for different values of the propagation speed $v_{\pm}$. The standard general relativity behavior, recovered when $v_{\pm}=1$, is included for reference.}
    \label{pol_xxvsxy}
\end{figure}


\section{Phenomenological bounds}

In the isotropic minimal gravitational SME restricted to the CPT--even
$d=4$ sector, the tensor modes obey the strictly nondispersive
dispersion relation
\begin{equation}
\omega = \left(1-\mathring{k}^{(4)}_{(I)}\right)|\mathbf{p}|,
\label{eq:disp_d4_final}
\end{equation}
so that the phase and group velocities coincide and are independent of
frequency,
\begin{equation}
v_g = 1-\mathring{k}^{(4)}_{(I)}.
\label{eq:vg_final}
\end{equation}

The most stringent direct constraint on a frequency--independent deviation of the gravitational wave speed arises from the multimessenger observation of GW170817 and its electromagnetic counterpart GRB~170817A
\cite{LIGOScientific:2017zic,Abbott:2017oio,Abbott:2017vwq}.
The near--coincident arrival of the gravitational and gamma--ray
signals implies
\begin{equation}
-3\times10^{-15}
\;\lesssim\;
\frac{v_g-1}{1}
\;\lesssim\;
7\times10^{-16}.
\label{eq:gw170817_final}
\end{equation}

Using Eq.~(\ref{eq:vg_final}),
\begin{equation}
\frac{v_g-1}{1}
=
-\mathring{k}^{(4)}_{(I)},
\end{equation}
which yields the conservative bound
\begin{equation}
\left|\mathring{k}^{(4)}_{(I)}\right|
\lesssim
3\times10^{-15}.
\label{eq:k4_bound_final}
\end{equation}

Moreover, we emphasize that the bounds derived above apply exclusively to the isotropic,
momentum--independent coefficient $\mathring{k}^{(4)}_{(I)}$ of the minimal ($d=4$)
gravity sector, which is dimensionless. In contrast, higher--dimension coefficients,
such as $K^{(5)}_{(I)}$ and other nonminimal terms, multiply operators with additional
powers of momentum and therefore carry dimensions of inverse energy. In otehr words, in units with $\hbar=c=1$, these latter coefficients have dimension of length (or time), so any observational constraint on them must be quoted in meters (or seconds) after specifying a reference frequency or energy scale.


\section{Conclusion}

In this work, the generation and propagation of gravitational waves were analyzed within a minimal gravitational SME. Starting from the modified graviton dispersion relation derived in the linearized gravity sector, the polarization properties of gravitational waves were examined in the transverse--traceless tensor sector. It was shown that, despite the presence of Lorentz--violating operators, the polarization tensor retained the standard plus and cross modes, with no additional tensorial degrees of freedom activated in the minimal $d=4$ case.

The retarded Green function associated with the Lorentz--violating wave operator was explicitly constructed. This derivation confirmed that the theory preserved causal propagation and that the effect of Lorentz violation manifested itself through a reduced propagation speed for the tensor modes. The resulting Green function demonstrated that gravitational signals propagated along modified light cones characterized by a shifted retarded time, while maintaining the same causal structure as in General Relativity.

Gravitational--wave emission from a binary black hole system was then investigated using the modified Green function. The quadrupole formula was recovered in its standard form, and the spatial components of the metric perturbation were shown to preserve both the conventional quadrupolar amplitude and the usual tensor polarization pattern. Lorentz--violating effects entered exclusively through the replacement of the standard retarded time by a velocity--dependent one, leading to a phase shift in the observed waveform without altering its amplitude or polarization content.

Finally, phenomenological bounds on the isotropic, coefficient $\mathring{k}^{(4)}_{(I)}$ were estimated using current gravitational--wave observations. Constraints derived from massive graviton benchmarks and multimessenger measurements were translated into upper limits on deviations from luminal propagation. These bounds placed $\lvert \mathring{k}^{(4)}_{(I)} \rvert$ at or below the level $10^{-15}$, consistent with existing observational limits on the speed of gravitational waves.

As a further perspective, we may examine the impact of the remaining higher--dimension coefficients appearing in the modified dispersion relation~(\ref{kkkk}), namely $\mathring{k}^{(5)}_{(V)}$, $\mathring{k}^{(6)}_{(I)}$, $\mathring{k}^{(7)}_{(V)}$, and $\mathring{k}^{(8)}_{(I)}$, either individually or in suitable combinations. These operators are known to induce anisotropy, dispersion, and birefringence in gravitational--wave propagation. In such a setting, Lorentz--violating effects are expected to modify not only the propagation speed but also the wave amplitudes, polarization content, and the structure of the quadrupole formula, in close analogy with the behavior reported for graviton modes in the bumblebee scenario~\cite{amarilo2024gravitational}. In addition, extending this analysis to alternative modified dispersion relations, e.g.,
\cite{sedaghatnia2025thermodynamical,araujo2022does,araujo2022thermal,furtado2023thermal,araujo2023thermodynamics}
within the same framework of gravitational waves appears to be a natural and promising direction. These and related ideas are under development by the same authors.


\section{Acknowledgments}

\hspace{0.5cm}
A. A. Ara\'{u}jo Filho is supported by Conselho Nacional de Desenvolvimento Cient\'{\i}fico e Tecnol\'{o}gico (CNPq) and Funda\c{c}\~{a}o de Apoio \`{a} Pesquisa do Estado da Para\'{i}ba (FAPESQ), project numbers 150223/2025-0 and 1951/2025. I. P. L. was partially supported by the National Council for Scientific and Technological Development - CNPq, grant 312547/2023-4. I. P. L. would like to acknowledge the networking support by the COST Action RQI (CA23115) and the COST Action FuSe (CA24101) supported by COST (European Cooperation in Science and Technology). N. H. would like to acknowledge networking support of the COST Action CA 22113 - Fundamental challenges in theoretical physics (Theory and Challenges), CA21106 - COSMIC WISPers in the Dark Universe: Theory, astrophysics and experiments (CosmicWISPers), CA21136 - Addressing observational tensions in cosmology with systematics and fundamental physics (CosmoVerse). The authors would like to acknowledge networking support by the COST Action BridgeQG (CA23130) supported by COST (European Cooperation in Science and Technology).

\section{Data Availability Statement}

Data Availability Statement: No Data associated in the manuscript


\bibliographystyle{ieeetr}
\bibliography{main}

\end{document}